\documentclass[epj]{svjour}
\usepackage{graphics}
\usepackage{amssymb}
\begin{document}

\title{On exact singular wave functions for identical planar charged
particles in uniform perpendicular magnetic field}
\author{A.Ralko \and T.T.Truong}

\institute{Laboratoire de Physique Th\'eorique et Mod\'elisation,
Universit\'e de Cergy-Pontoise, F-95031, Cergy-Pontoise
Cedex, France}

\date{To be published in Eur. Phys. J. B 2002}

\abstract{
We discuss the occurrence and properties of exact
singular anyonic wave functions describing stationary states of
two identical charged particles moving on a plane and under the
influence of a perpendicular uniform magnetic field.
\PACS{
      {03.65.Ge}{Solutions of wave equations: bound states} \and
      {05.30.Pr}{Fractional statistics systems} \and
      {71.10.Pm}{Fermions in reduced dimensions} \and {73.43.-f}{Quantum Hall effects.}} }
\maketitle

\section{Introduction}
\label{sec:1}
The general principles of quantum mechanics requires that
localized stationary states of a system be represented by square
integrable functions so as to be consistent with the probabilistic
interpretation of Born. This is in particular the situation for
the motion in a potential well rising indefinitely at infinity and
at a singular point localized, say at the origin. Here we are
concerned with this condition in two space dimensions. Let us
suppose that the wave function $\Psi$ behaves as $r^s u(r)$ near
the origin where $u(r)$ is a well-behaved function of $r$, the
distance from the origin, for $r \to 0$. Then for $\int |\Psi|^2
d^2 r$, to be convergent in a neighborhood of the origin, one
should have $s > - 1$. There is however a second condition stating
that the origin as a singular point should not be a source nor
sink for matter \cite{1}. The outward flow of the probability
flux $\vec{J}$ through a small sphere of radius $r$ must vanish as
$r \to 0$:
\[
\oint \vec{J} \cdot d \vec{a}= - \frac{i \hbar}{2 m} r^{2 s+1}
\int_{0}^{2 \pi} \left(u^* \frac{\partial u}{\partial r}- u
\frac{\partial u^*}{\partial r} \right)d \theta \rightarrow 0
\label{eq1}
\]
$m$ being the particle mass. This puts a stronger bound on $s$:
 i.e. $s>-\frac{1}{2}$. However if $u$ is a real function this
 condition is always satisfied as it is in the case we consider now.

 In most soluble two-dimensional problems presenting cylindrical
symmetric quantization of angular momentum to values $\hbar l$
leads to the condition $s^2 = l^2$ or $s = \pm |l|$ equivalently.
For the wave function to be single-valued one imposes $l \in
\mathbb{Z}$ and the choice $s=-|l|$ is excluded since it is
inconsistent with $s>-1$. But with the discovery of anyonic
behavior for wave functions of identical planar particles by
Leinaas $\&$ Myrheim \cite{2}, $l$ as quantized relative angular
momentum of two particles is a priori a real number: $l \in
\mathbb{R} $. Then it is possible to consider the second option
$s=-|l|$, so long as $s>-1$. This possibility which has been
pointed out a decade ago by Grundberg et al \cite{3} in a general
context leads to singular anyonic wave functions.

In this short note, we present examples of such functions
occurring in the relative motion of two identical planar particles
of mass $m$ and charge $e$ in a perpendicular uniform magnetic
field $B$. First we show how to construct the wave functions in
terms of Biconfluent Heun Polynomials and how quantization
conditions are imposed. In the next section we discuss the
structure of the low-lying singular wave functions. Finally we
close the paper with some comments and physical interpretations on
this unusual behavior.

\section{Construction of the wave functions}
\label{sec:2}
As the Coulomb repulsion between the particles has
rotational symmetry, we use cylindrical coordinates and the
symmetric gauge to describe the magnetic field. The Hamiltonian of
this relative motion reads:
\begin{eqnarray}
H = \frac{1}{m} \left[p_{r}^{2}+\left(\frac{p_{\theta}}{r}+
\frac{e B}{2}r \right)^2 \right]+
\kappa \frac{e^2}{r} \label{eq2}
\end{eqnarray}

Here, $r$ is the particle separation, $\theta$ its inclination
with respect to a reference axis, $p_{r}$ and $p_{\theta}$ are the
conjugate momenta and $\kappa = \frac{1}{4 \pi \epsilon_0}$ in
S.I. units.

The wave functions of stationary states $\Psi (r,\theta)$ are
solutions of the Schr\"odinger equation $H \Psi = E \Psi$ with
energy $E$. As $p_{\theta}$ is conserved and in view of possible
exotic statistics in two dimensions we may set $\Psi (r,\theta )
\simeq e^{i l \theta} R_{l}(r)$ with $l \in \mathbb{R}$. To put forward
the competition between magnetic length $l_{B}$ and the Bohr
radius $l_0$ defined by:
\begin{eqnarray}
l_{B}^{2}=\frac{2 \hbar}{e B} \ \ \ \ \ \ \ \textrm{and} \ \ \ \ \ \ \ \
l_{0}=\frac{\hbar^2}{\kappa e^2 m} \label{eq3}
\end{eqnarray}
we recast the radial Schr\"odinger equation for $R_{l}(r)$ in its dimensionless form:
\begin{eqnarray}
R_{l}''+\frac{1}{\xi} R_{l}'- \left[ \frac{l^2}{\xi^2}+\frac{C}{\xi}-
\xi^2 -2 (\epsilon-l)\right] R_{l}=0 \label{eq4}
\end{eqnarray}
using the scaled variables $\xi$, $\epsilon$ and $C$ given by:
\begin{eqnarray}
E=\frac{\hbar^2}{m l_{B}^2} \epsilon \ \ \ \ \ \ \ \ \ r= \xi l_{B}
\sqrt{2} \ \ \ \ \ \ \ \ \ \ C=\frac{l_{B}}{l_{0}} \sqrt{2}
\label{eq5}
\end{eqnarray}

>From eq.(\ref{eq4}), it can be seen that the asymptotic form of
$R_{l}(\xi)$ is $e^{\frac{- \xi^2}{2}}$. We set
$R_{l}(\xi)=v_{l}(\xi) e^{\frac{- \xi^2}{2}}$ with the assumption
that $v_{l}(\xi)$ is of milder growth, at $\xi \to \infty$, than
$e^{\frac{- \xi^2}{2}}$. $v_{l}(\xi)$ fulfills now an
``associated'' Biconfluent Heun equation:
\begin{eqnarray}
\xi v_{l}'' + (1-2 \xi^2) v_{l}' - \{ C + 2 (\epsilon -l-1) \xi \} v_{l} -
\frac{l^2}{\xi} v_{l}=0
\label{eq6}
\end{eqnarray}
with Maroni's canonical parameters $\alpha = \beta = 0$, \linebreak $\gamma =
2 (\epsilon-l)$ and $\delta = 2 C$ \cite{4}. We seek now a power
series solution of the form $v_{l}(\xi) = \sum_{n=0}^{\infty}
\alpha_{n} \xi^{s+n}$. Substitution of this series into
eq.(\ref{eq6}) yields the recursion relations:
\begin{eqnarray}
&&\lefteqn{ (s^2-l^2) \alpha_0 =0}, \nonumber \\
&&\left[ (s+1)^2-l^2 \right] \alpha_1 =C \alpha_0 ,\nonumber \\
&&\vdots \nonumber \\
&&\left[ (s+n)^2-l^2 \right] \alpha_n = \nonumber \\
&&C \alpha_{n-1} + 2[n+s-\epsilon+l-1] \alpha_{n-2}. \label{eq7}
\end{eqnarray}

Assuming $\alpha_{0} \neq 0$, we have $s^2 = l^2$ or $s=\sigma |l|$,
with \linebreak $\sigma=\pm$ . Now posing $\alpha_{n} = A_{n}^{(\sigma)}
\frac{1}{(1+2 s)_n n!}$ we recover the canonical recursion of the
Biconfluent Heun Function \linebreak $N(2 s,0,2 (\epsilon-l),2 C;\xi)$
\cite{4}:
\begin{eqnarray}
&&\lefteqn{ A_{n+2}^{(\sigma)} = C A_{n+1}^{(\sigma)}} \nonumber \\
&&- 2 (n+1)(n+1+2 s)(\epsilon-n-1-l-s) A_n^{(\sigma)}, \label{eq8}
\end{eqnarray}
with $A_0^{(\sigma)} = 1$ and $A_1^{(\sigma)} = C$; the parameter
$s$ may be $+ |l|$ or $- |l|$ and label regular or singular wave
functions near the origin.

As $N(2 s,0,2 (\epsilon-l),2 C;\xi)$ grows as $ e^{\xi^2}$
\cite{4} it is necessary to cut its defining power series down to
a polynomial by imposing two conditions on the recursion relation
(\ref{eq8}):
\begin{itemize}
\item energy quantization $\epsilon = \epsilon_n = (n+1+l+s)$,
\item and  $A_{n+1}^{(\sigma)}(C,l)=0$.
\end{itemize}

In general for given $B$ (or $C$) the angular momentum $l=l_{k}$
is the $k^{th}$ solution of an algebraic equation of order
$n$ for $n$ odd and of order $(n-1)$ for $n$ even. This has been already
pointed out in \cite{5}. The main consequence is that the
magnetic field fixes $l$ and thus gives rise to the expected
anyonic behavior. Note that the effective radial potential becomes
also modified for each eigenstate since it depends on $l$.

Globally the wave function of a stationary state reads:
\begin{eqnarray}
\psi^{(\sigma)}_{n,l_{k}}(\theta,\xi) \simeq e^{i l_{k} \theta}
 \xi^{\sigma |l_{k}|} \Pi_{n}(\xi,C) e^{-\frac{\xi^{2}}{2}} \label{eq9}
\end{eqnarray}
where $\Pi_{n}(\xi,C)$ is a special Biconfluent Heun polynomial,
obtained by truncating the power series.

As well-behaved wave functions with $\sigma = +$ have been already
discussed in \cite{5}, we shall be concerned here with the low-
lying $\psi^{(-)}_{n,l_{k}}(\theta,\xi)$ with $|l_{k}|<1$. The
corresponding energy eigenvalues are $
\epsilon^{(-)}_{n,l_{k}}=n+1+l_{k}-|l_{k}|$. In the following we shall
assume that $C\neq0$ (no infinitely strong magnetic field).

\section{Properties of the wave functions}
\label{sec:3}

They appear to be similar for pairs of the quantum number $n$.
This is essentially due to the fact that the conditions
$A_n^{(\sigma)}=0$ as well as $A_{n+1}^{(\sigma)}=0$ for $n$ even
lead to a pair of polynomials of same order in $C^{2}$ since
$A_{n+1}^{(\sigma)}$ contains an overall factor $C$. For
$\sigma=-$, we shall study the variation of $C^{2}$ as a function
of $l$ in the interval $-1<l<1$. Two possible behaviors of the
radial probability density for the relative separation $r$ arise:
divergent for $\frac{1}{2}<|l|<1$ and convergent for
$0<|l|<\frac{1}{2}$ in the neighborhood of zero. This is the only
physical relevant quantity since rotational symmetry yields a
uniform probability density for the inclination angle $\theta$.

\begin{figure}[ht]
\resizebox{0.5\textwidth}{!}{
\includegraphics{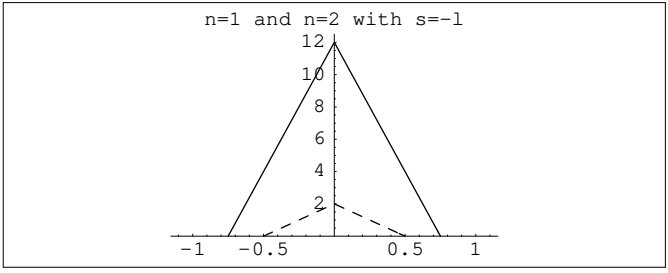}
} \caption{$C^2$ versus $l$ for $n=1$ (broken line) and $n=2$
(continuous line)} \label{fig:1}
\end{figure}

\begin{figure}[ht]
\resizebox{0.5\textwidth}{!}{
\includegraphics{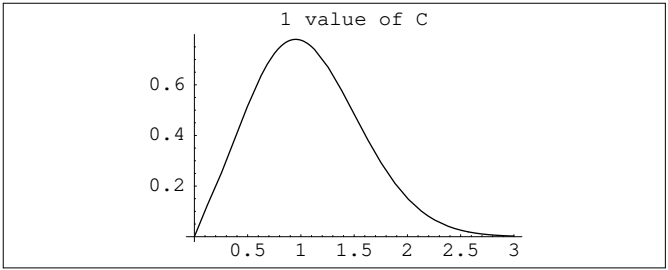}
} \caption{Probability distribution $P_{1}^{(-)}(r,C=1)$ for $n=1$}
\label{fig:2}
\end{figure}

\begin{figure}[ht]
\resizebox{0.5\textwidth}{!}{
\includegraphics{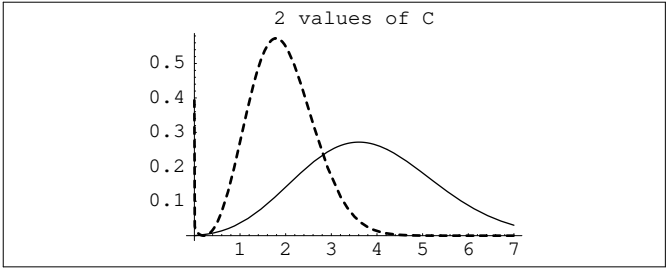}
} \caption{Probability distributions $P_{2}^{(-)}(r,C)$ for $n=2$, with $C=2 \sqrt{2}$ (continuous line) and  $C= \sqrt{2}$ (broken line).}
\label{fig:3}
\end{figure}

The first group of states is related to $n=1, 2$. The vanishing of
$A_{n+1}^{(\sigma)}=0$ yields linear relations between $C^{2}$ and
$l$, For $n=1,2$, the equations are respectively $
C^{2}=2(1-2|l|)$ and $C^{2}= 4(3-4|l|)$. This is shown in
Fig.\ref{fig:1}. We have given the probability distribution of
states $ r|\psi^{-}_{n,l_{k}}(\theta,r)|^{2}$ in Fig.\ref{fig:2}
and Fig.\ref{fig:3} as function of $r$.
\begin{eqnarray}
\lefteqn{P_{1}^{(-)}(r,C) \simeq \left( \frac{r}{l_B \sqrt{2}}\right)^{\frac{C^
2}{2}} \left(1 + \frac{2}{C} \frac{r}{l_B \sqrt{2}}\right)^{2} e^{-\frac{r^2}{2 l_{B}^{2}}}} \nonumber \\
&&P_{2}^{(-)}(r,C) \simeq \left( \frac{r}{l_B \sqrt{2}}\right)^{\frac{C^
2}{8}-\frac{1}{2}} \times \nonumber \\
&&\left(1 + \frac{8 r}{\sqrt{2} l_B (C^2-4)}+ \frac{2 r^2}{2 l_{B}^{2} (C^2-4)}\right)^{2} e^{-\frac{r^2}{2 l_{B}^{2}}} \label{eq10}
\end{eqnarray}
To save space, normalization factors with involved expressions are
not given but are implicitly taken into account. In particular, we
see that at some value of $C < 2$ for $n=2$ the radial probability
distribution starts to develop an integrable singularity at the
origin.

\begin{figure}[ht]
\resizebox{0.5\textwidth}{!}{
  \includegraphics{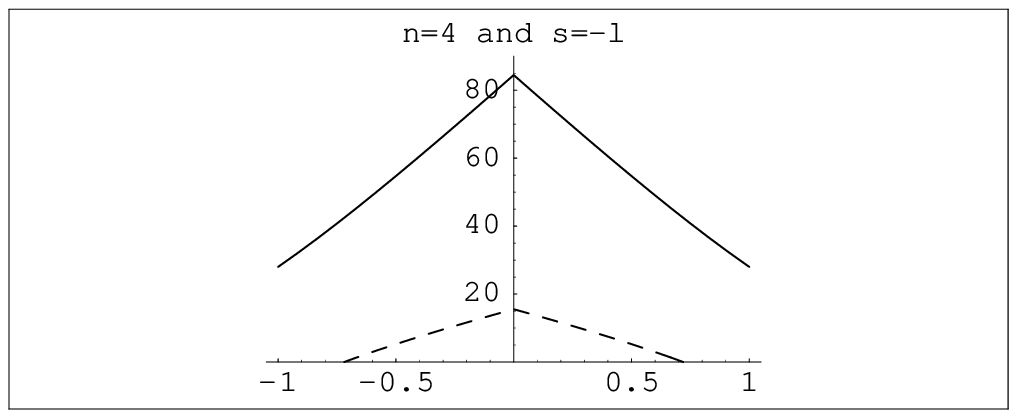}
}
\caption{$C^2$ versus of $l$ respectively for $n=4$, the upper (resp. lower) branch correspond to $+$ (resp. $-$) in expression of $C^2$}
\label{fig:6}
\end{figure}

The next group of states is given by $n=3,4$. The conditions
fixing $l$ in terms of $C$ are $A_{4}^{(-)}(C,l)=0$ and
$A_{5}^{(-)}(C,l)=0$, which correspond to the equations $C^{2}=
20(1-|l|) \pm \sqrt{64|l|^{2}-28|l|+73}$ and $C^{2}=(50-40|l|) \pm
6 \sqrt{16|l|^{2}-40|l|+33}$. Since $C=0$ is discarded we obtain
quadratic equations which yield two values $l_k$, $k=1,2$. In
Fig.\ref{fig:4},(resp. \ref{fig:6}) we plot $C^2$ in terms of $l$
for $n=3,(resp.4)$:
\begin{figure}[ht]
\resizebox{0.5\textwidth}{!}{
\includegraphics{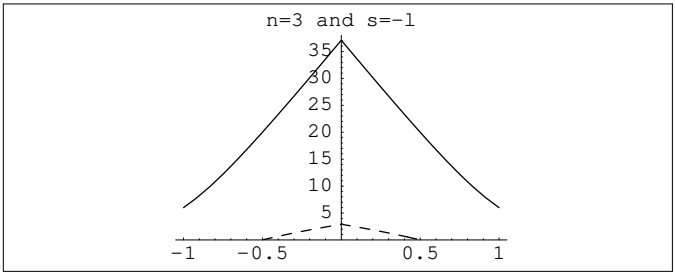}}
\caption{$C^2$ versus of $l$ respectively for $n=3$, the upper (resp. lower) branch correspond to $+$ (resp. $-$) in expression of $C^2$.}
\label{fig:4}
\end{figure}
\begin{figure}[ht]
\resizebox{0.5\textwidth}{!}{
\includegraphics{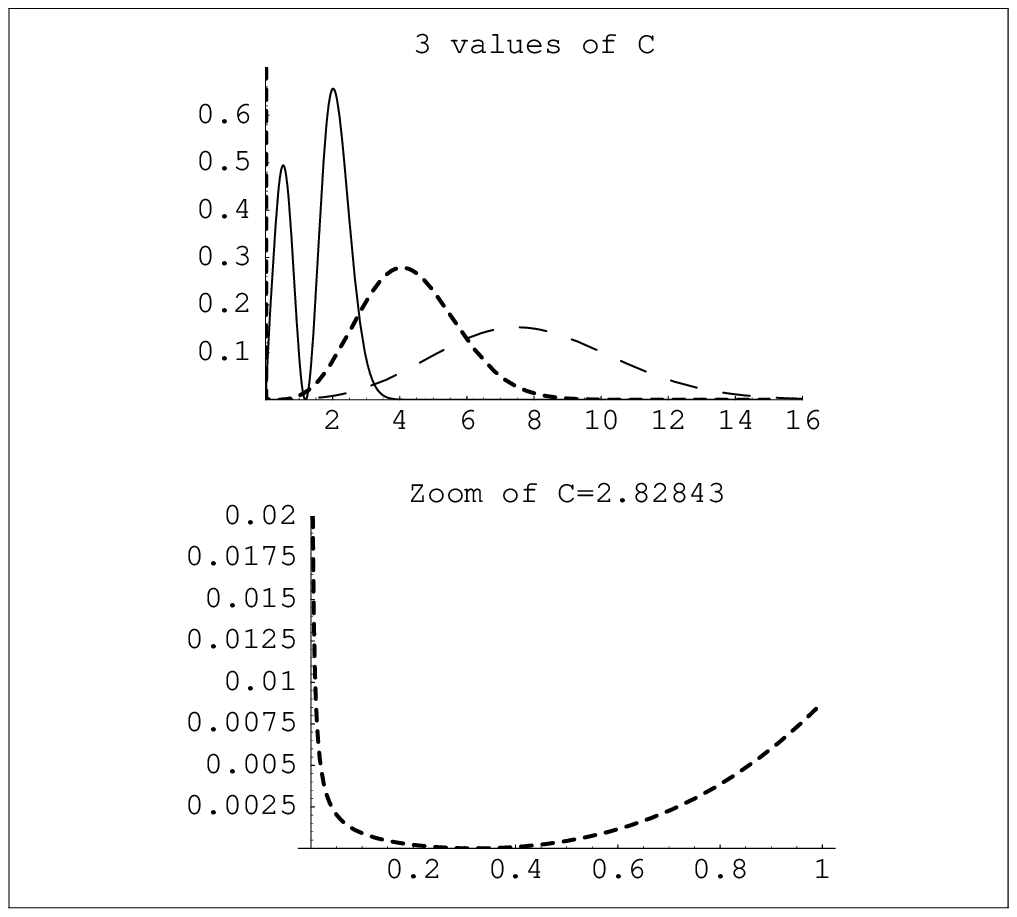}
} \caption{Probability distributions $P_{3}^{(-)}(r,C)$ for $n=3$}
\label{fig:5}
\end{figure}
\begin{figure}[ht]
\resizebox{0.5\textwidth}{!}{
\includegraphics{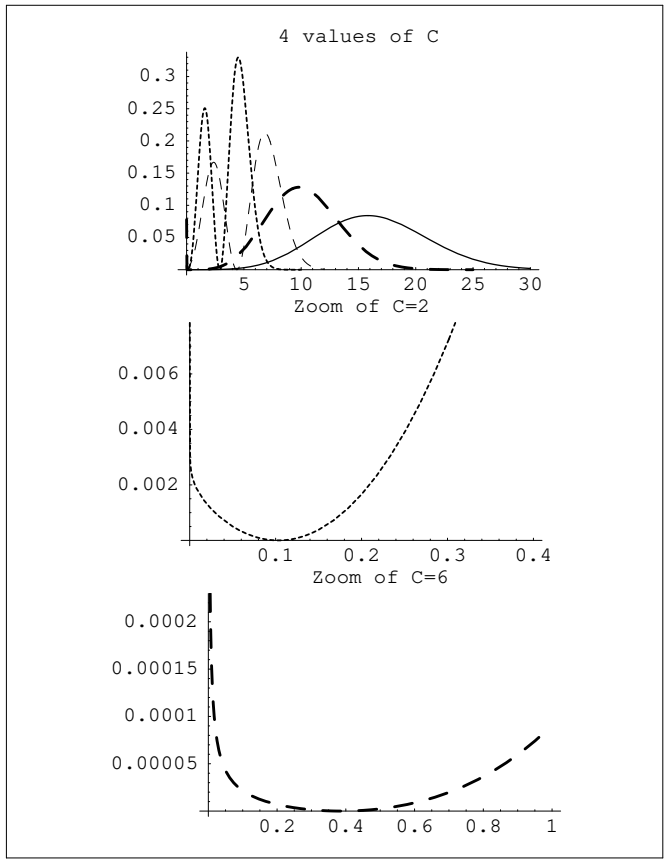}
} \caption{Probability distributions $P_{4}^{(-)}(r,C)$ for $n=4$}
\label{fig:7}
\end{figure}

We observe that for certain values of the magnetic field such that
$C^2$ is in a gap:
\begin{itemize}
\item $(20 - 2 \sqrt{73}) < C^2 < 6$  \ \ \ \ for $n=3$
\item $(50 - 6 \sqrt{33}) < C^2 < 28$  \ \ \ \ for $n=4$
\end{itemize}
there can be no states $\Psi_{n,l_k}^{(-)}(\theta,r)$ for $n=3,4$.
The behavior of the probability distributions depends on whether
$|l|$ is larger or smaller than $\frac{1}{2}$.

 For $n=3$ we may have 3 types of curves (see Fig.\ref{fig:5}):
\begin{itemize}
\item$P_{3}^{(-)}(r,C=5)$ (broken line) and
 $P_{3}^{(-)}(r,C=1)$ (continuous line) have smooth behavior at the origin.
\item $P_{3}^{(-)}(r,C=2 \sqrt{2})$ (dotted curve, with magnified part
in lower inset) is divergent but integrable at the origin.
\end{itemize}

For $n=4$, we present 2 pairs of curves in Fig.\ref{fig:7}:
\begin{itemize}
\item $P_{4}^{(-)}(r,C=9)$ (continuous line) and
$P_{4}^{(-)}(r,C=3)$ (thin broken line) which go smoothly to zero
at the origin
\item $P_{4}^{(-)}(r,C=6)$ (bold broken line) and $P_{4}^{(-)}(r,C=2)$
(dotted line) which diverge at $r=0$ but remain integrable in this
neighborhood. This curves are magnified in insets in the lower
part of Fig.\ref{fig:7}.
\end{itemize}

The $P_{3,4}^{(-)}(r,C)$ are computed with the expression of the
wave function in eq.(\ref{eq9}). Since the form of the special
Biconfluent Heun Polynomials $\Pi_{3,4}(\xi,C)$ are quite
involved, we shall not give here the explicit form of
$P_{3,4}^{(-)}(r,C)$.

Higher stationary states come also in pairs $n=5,6$ and $n=7,8$
with algebraic relations of third and fourth order in $C^2$
(and/or $l$). The angular momentum $l$ will have respectively
three and four values and a more complicated structure in terms of
magnetic field arises. General features concerning two classes of
smooth and singular behavior at the origin however remain. We
shall discuss them in a more extensive study later on. Beyond
these values of $n$, one can only tackle the problem numerically
since the algebraic equations of higher order do not posses
solutions in close form.

\begin{table}[ht]
\begin{center}
\begin{tabular}{|c|c|c|c|c|c|}
\hline
 & & & & & \\
$n$ & $B$ & $C$ & $\frac{r_0}{l_0}$ & $\epsilon_{n,l_k}^{(-)}$ & $\frac{r_1}{l_0}$ \\
 & & & & & \\ \hline
$1$ & $940215$ & $1$ & $0$ & $3$ & $0.3409025$\\ \hline
$2$ & $470108$ & $\sqrt{2}$ & $0.207107$ & $1.75$ & $0.5$\\
\cline{2-6}
 & $117527$ & $2 \sqrt{2} $ & $0$ & $0.625$ & $0.380453$ \\ \hline
$3$ & $940215$ & $1$ & $0$ & $6.60434$ & $0.215634$\\
\cline{2-6}
 & $37608.6$ & $5$ & $0$ & $0.263969$ & $3.84573$ \\
\cline{2-6}
 & $117527$ & $2 \sqrt{2} $ & $0.322087$ & $0.545623$ & $1.99029$\\ \hline
$4$ & $235054$ & $2$ & $0.1032474$ & $1.9443$ & $0.666414$\\
\cline{2-6}
 & $104468$ & $3$ & $0$ & $0.965468$ & $1.07414$ \\
\cline{2-6}
 & $26117.1$ & $6$ & $0.386699$ & $0.184408$ & $5.4233$\\
\cline{2-6}
 & $11607.6$ & $9$ & $0$ & $0.120639$ & $9.1545$ \\ \hline
\end{tabular}
\end{center}
\caption{ Table of closest zeros $r_0$ of low-lying probability
distributions with given $C$, energy levels and the locations of
the classically limiting points $r_1$ at these values of the
energy ($\epsilon_{n,l_k}^{(-)}=n+1-2 |l|$). Values of the
magnetic $B$ in Tesla, corresponding to chosen values of $C$ are
listed in the second column.}
\label{tab:1}
\end{table}

\section{Comments and conclusion}

The presence of fractional power singularities of the wave
functions at the origin of coordinates calls for some comments.
The rise of the probability of finding a separation $r$ between
two identical particles within $dr$ suggests a tendency for the
particles to cluster together. Since \linebreak $P_{n}^{(-)}(r,C)
dr \simeq r^{-\lambda} dr \simeq d \left( r^{1-\lambda} \right)$
and as $0<\lambda<1$, integration of $r$ from $0$ to any value
$r=a$ close to the origin is finite. This means that the
probability of finding any $r$ in the interval $[0,a]$ is always
finite, despite the apparent divergence. Moreover the clustering
tendency in the presence of classical Coulomb repulsion is
somewhat counter-intuitive. To reconcile this behavior with
quantum mechanics we have given in Table. \ref{tab:1} the values
of the zeros $r_0$ of the probability distributions nearest to
zero, from which start divergent behaviors for various $n$ and
$l_k$ as well as the locations $r_1$ below which classical motion
is not possible in corresponding effective potentials. For the
four low lying $n$, we see that always $r_0 < r_1$. Hence the
divergence occurs only in forbidden classical regions of the
motion, thus not directly observable. Anyhow, such situations
occur also at tremendously high $B$-field, of the order of $10^5$
Tesla.

Finally, it is amusing to note that there is an analog of this
situation in electrostatics. The electric field, near the edge of
a two-dimensional wedge of opening angle $\theta$ is the analog of
$\psi^{(\sigma)}_{n,l_{k}}(\theta,r)$ and behaves as
$r^{\frac{\pi}{\theta}-1}$, $r$ being the distance from the edge
\cite{6}. Thus it diverges for \linebreak $\pi<\theta<2\pi$. Now the electric
field energy density is the analog of $P^{-}_{n}(r,C)$, but here
it has always a nice behavior $r^{2\frac{\pi}{\theta}-1}$ for
$0<\theta<2\pi$.


\begin{thebibliography}{}
\bibitem{1}
L.E.Ballentine, \textit{Quantum Mechanics: a modern
development} (World Scientific, Singapore, 1998) pp. 106.
\bibitem{2}
J.M.Leinaas $\&$ J.Myrheim, Nuovo Cimento \textbf{B 37}, (1997) 1.
\bibitem{3}
J.Grundberg, T.H.Hansson, A.Karhhede $\&$ J.M.Leinaas, Mod. Phys. Lett.
\textbf{B 5}, (1991) 539.
\bibitem{4}
P.Maroni, \textit{Heun's Differential Equations}
(Oxford Univ. Press, New York 1995) pp. 191. \\
F.Batola, \textit{Th\`ese de troisi\`eme cycle}, Univ. de Paris 6
\bibitem{5}
T.T.Truong $\&$ D.Bazzali, Phys. Lett. A \textbf{269}, (2000) 186.
\bibitem{6} J.D.Jackson, \textit{Classical Electrodynamics} (John Wiley, New.York, 1999) pp.75.
\end{thebibliography}
\end{document}